\documentclass[fleqn,10pt,dvipsnames]{wlscirep}
\usepackage[utf8]{inputenc}
\usepackage[T1]{fontenc}

\title{Warwick Electron Microscopy Datasets}

\author[1,*]{Jeffrey M. Ede}
\affil[1]{University of Warwick, Department of Physics, Coventry, CV4 7AL, UK}
\affil[*]{j.m.ede@warwick.ac.uk}

\newcommand*\linkcolours{CornflowerBlue}

\usepackage{times}
\usepackage{graphicx}
\usepackage{amssymb}
\usepackage{gensymb}
\usepackage{amsmath}
\usepackage{breakurl}

\usepackage{url,hyperref}
\hypersetup{
colorlinks,
linkcolor=\linkcolours,
citecolor=\linkcolours,
filecolor=\linkcolours,
urlcolor=\linkcolours
}

\usepackage{mathtools, cuted}

\usepackage{tabularx}
\usepackage{multirow}

\usepackage{float}

\newcommand*\myimagescale{1.3}

\newcolumntype{Y}{>{\centering\arraybackslash}X}

\usepackage[]{placeins}

\newcommand\extravspace{\vspace{3pt}}

\usepackage{pgf}

\usepackage[percent]{overpic}

\begin{abstract}
Large, carefully partitioned datasets are essential to train neural networks and standardize performance benchmarks. As a result, we have set up new repositories to make our electron microscopy datasets available to the wider community. There are three main datasets containing 19769 scanning transmission electron micrographs, 17266 transmission electron micrographs, and 98340 simulated exit wavefunctions, and multiple variants of each dataset for different applications. To visualize image datasets, we trained variational autoencoders to encode data as 64-dimensional multivariate normal distributions, which we cluster in two dimensions by t-distributed stochastic neighbor embedding. In addition, we have improved dataset visualization with variational autoencoders by introducing encoding normalization and regularization, adding an image gradient loss, and extending t-distributed stochastic neighbor embedding to account for encoded standard deviations. Our datasets, source code, pretrained models, and interactive visualizations are openly available at \url{https://github.com/Jeffrey-Ede/datasets}.
\end{abstract}

\begin{document}

\flushbottom
\maketitle
\thispagestyle{empty}

\section{Introduction}

We have set up new repositories\cite{warwickem!} to make our large new electron microscopy datasets available to both electron microscopists and the wider community. There are three main datasets containing 19769 experimental scanning transmission electron microscopy\cite{fei_intro} (STEM) images, 17266 experimental transmission electron microscopy\cite{fei_intro} (TEM) images and 98340 simulated TEM exit wavefunctions\cite{preprint+ede2020exit}. Experimental datasets represent general research and were collected by dozens of University of Warwick scientists working on hundreds of projects between January 2010 and June 2018. We have been using our datasets to train artificial neural networks (ANNs) for electron microscopy\cite{ede2019improving, ede2020partial, preprint+ede2020exit, ede2020adaptive, preprint+ede2019deep}, where standardizing results with common test sets has been essential for comparison. This paper provides details of and visualizations for datasets and their variants, and is supplemented by source code, pretrained models, and both static and interactive visualizations\cite{datasets_repo}.

Machine learning is increasingly being applied to materials science\cite{schmidt2019recent, von2020introducing}, including to electron microscopy\cite{belianinov2015big}. Encouraging scientists, ANNs are universal approximators\cite{hornik1989multilayer} that can leverage an understanding of physics to represent\cite{lin2017does} the best way to perform a task with arbitrary accuracy. In theory, this means that ANNs can always match or surpass the performance of contemporary methods. However, training, validating and testing requires large, carefully partitioned datasets\cite{raschka2018model, roh2019survey} to ensure that ANNs are robust to general use. To this end, our datasets are partitioned so that each subset has different characteristics. For example, TEM or STEM images can be partitioned so that subsets are collected by different scientists, and simulated exit wavefunction partitions can correspond to Crystallography Information Files\cite{hall1991crystallographic} (CIFs) for materials published in different journals.

Most areas of science are facing a reproducibility crisis\cite{baker2016reproducibility}, including artificial intelligence\cite{hutson2018artificial}. Adding to this crisis, natural scientists do not always benchmark ANNs against standardized public domain test sets; making results difficult or impossible to compare. In electron microscopy, we believe this is a symptom of most datasets being small, esoteric or not having default partitions for machine learning. For example, most datasets in the Electron Microscopy Public Image Archive\cite{iudin2016empiar, hey2020machine} are for specific materials and are not partitioned. In contrast, standard machine learning datasets such as CIFAR-10\cite{krizhevsky2014cifar, krizhevsky2009learning}, MNIST\cite{lecun2010mnist}, and ImageNet\cite{russakovsky2015imagenet} have default partitions for machine learning and contain tens of thousands or millions of examples. By publishing our large, carefully partitioned machine learning datasets, and setting an example by using them to standardize our research, we aim to encourage higher standardization of machine learning research in the electron microscopy community. 

There are many popular algorithms for high-dimensional data visualization\cite{tenenbaum2000global, roweis2000nonlinear, zhang2007mlle, donoho2003hessian, belkin2003laplacian, zhang2004principal, buja2008data, van2014accelerating} that can map $N$ high-dimensional vectors of features $\{ \textbf{x}_1, ..., \textbf{x}_N \},\; \textbf{x}_i \in \mathbb{R}^u$ to low-dimensional vectors $\{ \textbf{y}_1, ..., \textbf{y}_N \},\; \textbf{y}_i \in \mathbb{R}^v$. A standard approach for data clustering in $v \in \{ 1,2,3 \}$ dimensions is t-distributed stochastic neighbor embedding\cite{maaten2008visualizing, wattenberg2016use} (tSNE). To embed data by tSNE, Kullback-Leibler (KL) divergence,
\begin{equation}
    L_\text{tSNE} = \sum\limits_i \sum\limits_{j \neq i} p_{ij} \log\left( \frac{p_{ij}}{q_{ij}} \right)\,,
\end{equation}
is minimized by gradient descent\cite{ruder2016overview} for normally distributed pairwise similarities in real space, $p_{ij}$, and heavy-tailed Student t-distributed pairwise similarities in an embedding space, $q_{ij}$. For symmetric tSNE\cite{maaten2008visualizing},
\begin{align}
    p_{i|j} &= \frac{\exp\left(-||\textbf{x}_i-\textbf{x}_j||_2^2 / 2\alpha_j^2\right)}{
    \sum\limits_{k\neq j} \exp\left(-||\textbf{x}_k-\textbf{x}_j||_2^2 / 2\alpha_j^2\right) } \label{eqn:traditional_pij}\,, \\
    p_{ij} &= \frac{p_{i|j} + p_{j|i}}{2N}\,, \\ 
    q_{ij} &= \frac{\left(1 + ||\textbf{y}_i-\textbf{y}_j||_2^2 \right)^{-1}}{
    \sum\limits_{k\neq i} \left(1 + ||\textbf{y}_k-\textbf{y}_i||_2^2 \right)^{-1} }\,.
\end{align}
To control how much tSNE clusters data, perplexities of $p_{i|j}$ for $j \in \{1, ..., N \}$ are adjusted to a user-provided value by fitting $\alpha_j$. Perplexity, $\exp(H)$, is an exponential function of entropy, $H$, and most tSNE visualizations are robust to moderate changes to its value. 

Feature extraction is often applied to decrease input dimensionality, typically to $u \lesssim 100$, before clustering data by tSNE. Decreasing input dimensionality can decrease data noise and computation for large datasets, and is necessary for some high-dimensional data as distances, $||\textbf{x}_i-\textbf{x}_j||_2$, used to compute $p_{ij}$ are affected by the curse of dimensionality\cite{schubert2017intrinsic}. For image data, a standard approach\cite{maaten2008visualizing} to extract features is probabilistic\cite{halko2011finding, mart2011arand} or singular value decomposition\cite{wall2003singular} (SVD) based principal component analysis\cite{jolliffe2016principal} (PCA). However, PCA is limited to linearly separable features. Other hand-crafted feature extraction methods include using a histogram of oriented gradients\cite{dalal2005histograms}, speeded-up robust features\cite{bay2008speeded}, local binary patterns\cite{ojala2002multiresolution}, wavelet decomposition\cite{mallat1989theory} and other methods\cite{latif2019content}. The best features to extract for a visualization depend on its purpose. However, most hand-crafted feature extraction methods must be tuned for different datasets. For example, Minka's algorithm\cite{minka2001automatic} is included in the scikit-learn\cite{van2014scikit} implementation of PCA by SVD to obtain optimal numbers of principal components to use.  

To increase representation power, nonlinear and dataset-specific features can be extracted with deep learning. For example, by using the latent space of an autoencoder\cite{tschannen2018recent, kramer1991nonlinear} (AE) or features before logits in a classification ANN\cite{marcelino2018transfer}. Indeed, we have posted AEs for electron microscopy with pre-trained models\cite{kernels+MLPs+Autoencoders_repo, preprint+ede2018autoencoders} that could be improved. However, AE latent vectors can exhibit inhomogeneous dimensional characteristics and pathological semantics, limiting correlation between latent features and semantics. To encode well-behaved latent vectors suitable for clustering by tSNE, variational autoencoders\cite{kingma2014auto, kingma2019introduction} (VAEs) can be trained to encode data as multivariate probability distributions. For example, VAEs are often regularized to encode multivariate normal distributions by adding KL divergence of encodings from a standard normal distribution to its loss function\cite{kingma2014auto}. The regularization homogenizes dimensional characteristics and sampling noise correlates semantics with latent features.

\section{Dataset Visualization}\label{sec:visualization}

\begin{figure*}[tbp!]
\centering
\includegraphics[width=0.85\textwidth]{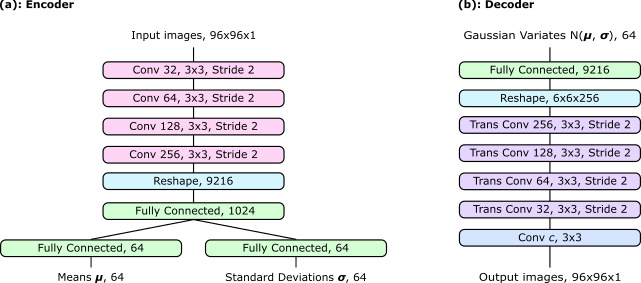}
\caption{ Simplified VAE architecture. a) An encoder outputs means, $\boldsymbol\mu$, and standard deviations, $\boldsymbol\sigma$, to parameterize multivariate normal distributions, $\textbf{z} \sim \text{N}(\boldsymbol\mu, \boldsymbol\sigma)$. b) A generator predicts input images from $\textbf{z}$. }
\label{vae_architecture}
\end{figure*}

To visualize datasets presented in this paper, we trained VAEs shown in figure~\ref{vae_architecture} to embed 96$\times$96 images in $u=64$ dimensions before clustering in $v=2$ dimensions by tSNE. Our VAE consists of two convolutional neural networks\cite{mccann2017convolutional, krizhevsky2012imagenet} (CNNs): an encoder and a generator. The encoder embeds batches of $B$ input images, $I$, as mean vectors, $\{\boldsymbol\mu_1, ..., \boldsymbol\mu_B\}$, and standard deviation vectors, $\{\boldsymbol\sigma_1, ..., \boldsymbol\sigma_B\}$, to parameterize multivariate normal distributions. During training, input images are linearly transformed to have minimum and maximum values of 0 and 1, respectively, and we apply a random combination of flips and 90$\degree$ rotations to augment training data by a factor of eight. The generator, $G$, is trained to cooperate with the encoder to output encoder inputs by sampling latent vectors, $\textbf{z}_i = \boldsymbol\mu_i + \boldsymbol\sigma_i \boldsymbol\epsilon_i$, where $\boldsymbol\mu_i = \{\mu_{i1},...,\mu_{iu}\}$, $\boldsymbol\sigma_i = \{\sigma_{i1},...,\sigma_{iu}\}$, and $\boldsymbol\epsilon_i = \{\epsilon_{i1},...,\epsilon_{iu}\}$ are random variates sampled from standard normal distributions, $\epsilon_{ij} \sim N(0, 1)$. Each convolutional or fully connected layer is followed by batch normalization\cite{ioffe2015batch} then ReLU\cite{nair2010rectified} activation, except the output layers of the encoder and generator. An absolute nonlinearity, $f(x) = |x|$, is applied to encode positive standard deviations.

\begin{figure*}[tbp!]
\centering
\includegraphics[width=\textwidth]{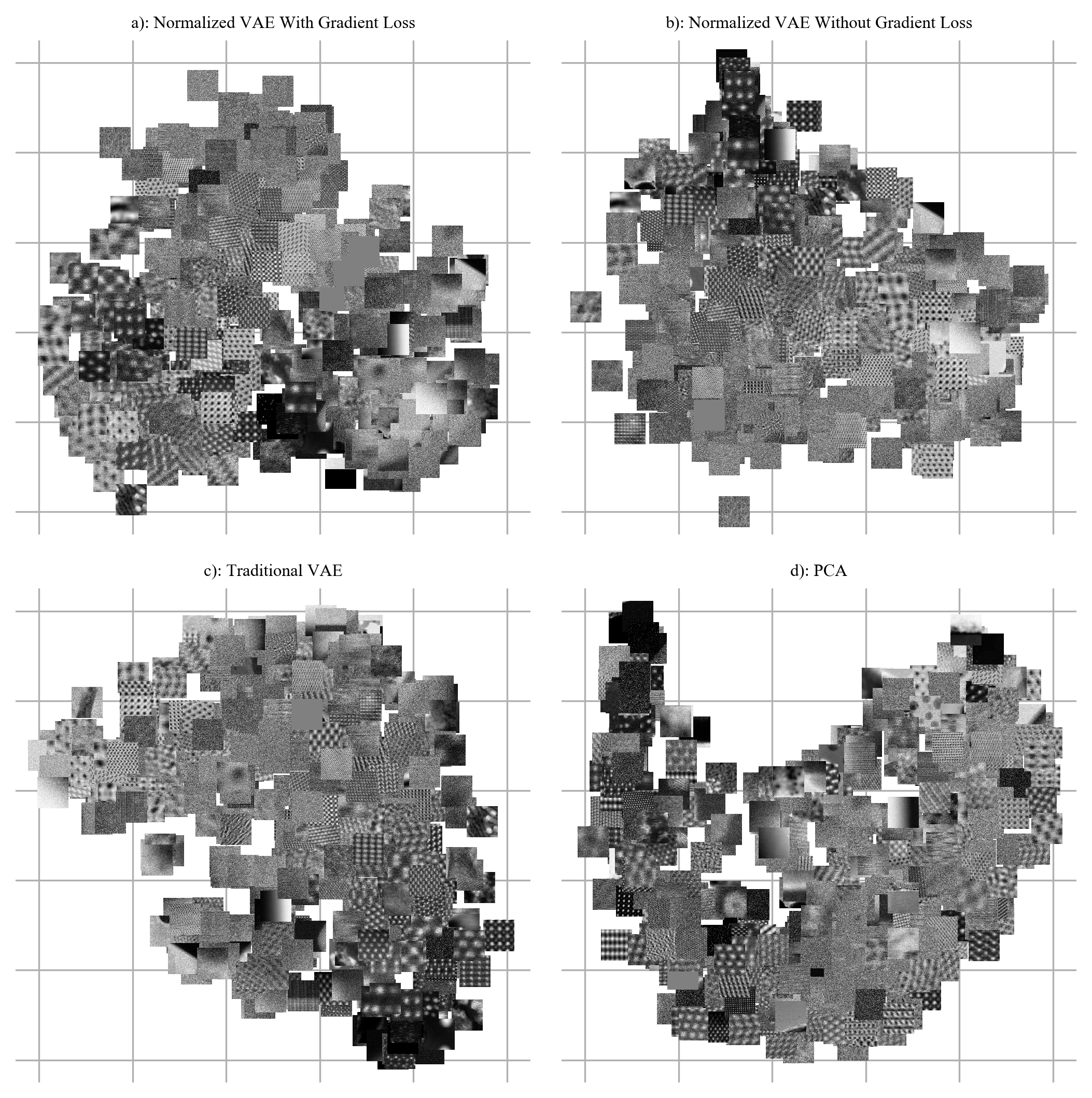}
\caption{ Images at 500 randomly selected images in two-dimensional tSNE visualizations of 19769 96$\times$96 crops from STEM images for various embedding methods. Clustering is best in a) and gets worse in order a)$\rightarrow$b)$\rightarrow$c)$\rightarrow$d). }
\label{four_panel_tsne}
\end{figure*}

Traditional VAEs are trained to optimize a balance, $\lambda_\text{MSE}$, between mean squared errors (MSEs) of generated images and KL divergence of encodings from a multivariate standard normal distribution\cite{kingma2014auto},
\begin{equation}\label{eqn:traditional_vae}
    L_\text{trad} = \lambda_\text{MSE} \text{MSE}( G(\textbf{z}), I ) + \frac{1}{2Bu} \sum\limits_{i=1}^B \sum\limits_{j=1}^u \mu_{ij}^2 + \sigma_{ij}^2 - \log(\sigma_{ij}^2) - 1.
\end{equation}
However, traditional VAE training is sensitive to $\lambda_\text{MSE}$\cite{higgins2017beta} and other hyperparameters\cite{hu2019parameter}. If $\lambda_\text{MSE}$ is too low, the encoder will learn learn to consistently output $\sigma_{ij} \simeq 1$, limiting regularization. Else if $\lambda_\text{MSE}$ is too high, the encoder will learn to output $\sigma_{ij} \ll |\mu_{ij}|$, limiting regularization. As a result, traditional VAE hyperparameters must be carefully tuned for different ANN architectures and datasets. To improve VAE regularization and robustness to different datasets, we normalize encodings parameterizing normal distributions to
\begin{align}
    \mu_{ij} &\leftarrow \frac{\lambda_\mu (\mu_{ij} - \mu_{\text{avg},j})}{\mu_{\text{std},j}}\,, \\
    \sigma_{ij} &\leftarrow \frac{\sigma_{ij}}{2\sigma_{\text{std},j}}\,,
\end{align}
where batch means and standard deviations are
\begin{align}
    \mu_{\text{avg},j} &= \frac{1}{B}\sum\limits_{k=1}^B \mu_{kj}\,, \\
    \mu_{\text{std},j}^2 &= \frac{1}{B}\sum\limits_{k=1}^B \mu_{kj}^2 - \left(\frac{1}{B}\sum\limits_{k=1}^B \mu_{kj}\right)^2\,, \\
    \sigma_{\text{std},j}^2 &= \frac{1}{B}\sum\limits_{k=1}^B \sigma_{kj}^2 - \left(\frac{1}{B}\sum\limits_{k=1}^B \sigma_{kj}\right)^2\,.
\end{align}
Encoding normalization is a modified form of batch normalization\cite{ioffe2015batch} for VAE latent spaces. As part of encoding normalization, we introduce a new hyperparameter, $\lambda_\mu$, to scale the ratio of expectations $\text{E}(| \mu_{ij} |)/\text{E}(| \sigma_{ij} |)$. We use $\lambda_\mu = 2.5$ in this paper; however, we confirm that training is robust to values $\lambda_\mu \in \{1.0, 2.0, 2.5\}$ for a range of datasets and ANN architectures. Batch means are subtracted from $\boldsymbol\mu$ and not $\boldsymbol\sigma$ so that $\sigma_{ij} \ge 0$. In addition, we multiply $\sigma_{\text{std},j}$ by an arbitrary factor of 2 so that $\text{E}(| \mu_{ij} |) \approx \text{E}(| \sigma_{ij} |)$ for $\lambda_\mu = 1$.

Encoding normalization enables the KL divergence loss in equation~\ref{eqn:traditional_vae} to be removed as latent space regularization is built into the encoder architecture. However, we find that removing the KL loss can result in VAEs encoding either very low or very high $\sigma_{ij}$. In effect, an encoder can learn to use $\boldsymbol\sigma$ apply a binary mask to $\boldsymbol\mu$ if a generator learns that latent features with very high absolute values are not meaningful. To prevent extreme $\sigma_{ij}$, we add a new encoding regularization loss, $\text{MSE}(\boldsymbol\sigma, 1)$, to the encoder. Human vision is sensitive to edges\cite{mcilhagga2018estimates}, so we also add a gradient-based loss to improve realism. Adding a gradient-based loss is a computationally inexpensive alternative to training a variational autoencoder generative adversarial network\cite{larsen2015autoencoding} (VAE-GAN) and often achieves similar performance. Our total training loss is 
\begin{equation}\label{eqn:improved_vae}
    L = \lambda_\text{MSE} \text{MSE}( G(\textbf{z}), I ) + \lambda_\text{Sobel} \text{MSE}( S(G(\textbf{z})), S(I) ) + \text{MSE}(\boldsymbol\sigma, 1)\,,
\end{equation}
where we chose $\lambda_\text{MSE} = \lambda_\text{Sobel} = 50$, and $S(x)$ computes a concatenation of horizontal and vertical Sobel derivatives\cite{vairalkar2012edge} of $x$. We found that training is robust to choices of $\lambda_\text{MSE} = \lambda_\text{Sobel}$ where $\lambda_\text{MSE} \text{MSE}( G(\textbf{z}), I ) + \lambda_\text{Sobel} \text{MSE}( S(G(\textbf{z})), S(I) )$ is in $[0.5, 25.0]$, and have not investigated losses outside this interval.

We trained VAEs to minimize $L$ by ADAM\cite{kingma2014adam} optimized stochastic gradient descent\cite{ruder2016overview, zou2018stochastic}. At training iteration $t \in [1, T]$, we used a stepwise exponentially decayed learning rate\cite{ge2019step},  
\begin{equation}
    \eta = \eta_\text{start} a^{\text{floor} (bt/T)}\,,
\end{equation}
and a DEMON\cite{chen2019decaying} first moment of the momentum decay rate,
\begin{equation}
    \beta_1 = \frac{\beta_\text{start} (1 - t/T)}{(1-\beta_\text{start}) + \beta_\text{start}(1 - t/T)}\,,
\end{equation}
where we chose initial values $\eta_\text{start} = 0.001$ and $\beta_\text{start} = 0.9$, exponential base $a = 0.5$, $b=8$ steps, and $T=600000$ iterations. We used a batch size of $B=64$ and emphasize that a large batch size decreases complication of encoding normalization by varying batch statistics. Training our VAEs takes about 12 hours on a desktop computer with an Nvidia GTX 1080 Ti GPU and an Intel i7-6700 CPU.

To use VAE latent spaces to cluster data, means are often embedded by tSNE. However, this does not account for highly varying $\sigma$ used to calculate latent features. To account for uncertainty, we modify calculation of pairwise similarities, $p_{ij}$, in equation~\ref{eqn:traditional_pij} to include both $\boldsymbol\mu_i$ and $\boldsymbol\sigma_i$ encoded for every example, $i \in [1, N]$, in our datasets,
\begin{align}
    p_{i|j} &= \exp\left(- \frac{1}{2\alpha_j^2} \sum\limits_k w_{ijk} (\mu_{ik} - \mu_{jk})^2 \right) \left(
    \sum\limits_{m\neq j} \exp\left(- \frac{1}{2\alpha_j^2} \sum\limits_k w_{mjk} (\mu_{mk} - \mu_{jk})^2\right) \right)^{-1}\,, 
\end{align}
where we chose weights
\begin{align}
    w_{ijk} &= \frac{1}{\sigma_{ik}^2 + \sigma_{jk}^2 + \epsilon} \left( \sum\limits_l 
    \frac{1}{\sigma_{il}^2 + \sigma_{jl}^2 + \epsilon} \right)^{-1}\,.
\end{align}
We add $\epsilon = 0.01$ for numerical stability, and to account for uncertainty in $\boldsymbol\sigma$ due to encoder imperfections or variation in batch statistics. Following Oskolkov\cite{oskolkov2019how}, we fit $\alpha_j$ to perplexities given by $N^{1/2}$, where $N$ is the number of examples in a dataset, and confirm that changing perplexities by $\pm 100$ has little effect on visualizations for our $N \simeq 20000$ TEM and STEM datasets. To ensure convergence, we run tSNE computations for 10000 iterations. In comparison, KL divergence is stable by 5000 iterations for our datasets. In preliminary experiments, we observe that tSNE with $\boldsymbol\sigma$ results in comparable visualizations to tSNE without $\boldsymbol\sigma$, and think that tSNE with $\boldsymbol\sigma$ may be a slight improvement. For comparison, pairs of visualizations with and without $\boldsymbol\sigma$ are indicated in supplementary information.

Our improvements to dataset visualization by tSNE are showcased in figure~\ref{four_panel_tsne} for various embedding methods. The visualizations are for a new dataset containing 19769 96$\times$96 crops from STEM images, which will be introduced in section~\ref{sec:STEM}. To suppress high-frequency noise during training, images were blurred by a 5$\times$5 symmetric Gaussian kernel with a 2.5 px standard deviation. Clusters are most distinct in figure~\ref{four_panel_tsne}a for encoding normalized VAE training with a gradient loss described by equation~\ref{eqn:improved_vae}. Ablating the gradient loss in figure~\ref{four_panel_tsne}b results in similar clustering; however, the VAE struggles to separate images of noise and fine atom columns. In contrast, clusters are not clearly separated in figure~\ref{four_panel_tsne}c for a traditional VAE described by equation~\ref{eqn:traditional_vae}. Finally, embedding the first 50 principal components extracted by a scikit-learn\cite{scikit-learn} implementation of probabilistic PCA in figure~\ref{four_panel_tsne}d does not result in clear clustering.

\begin{figure*}[tbp!]
\centering
\includegraphics[width=\textwidth]{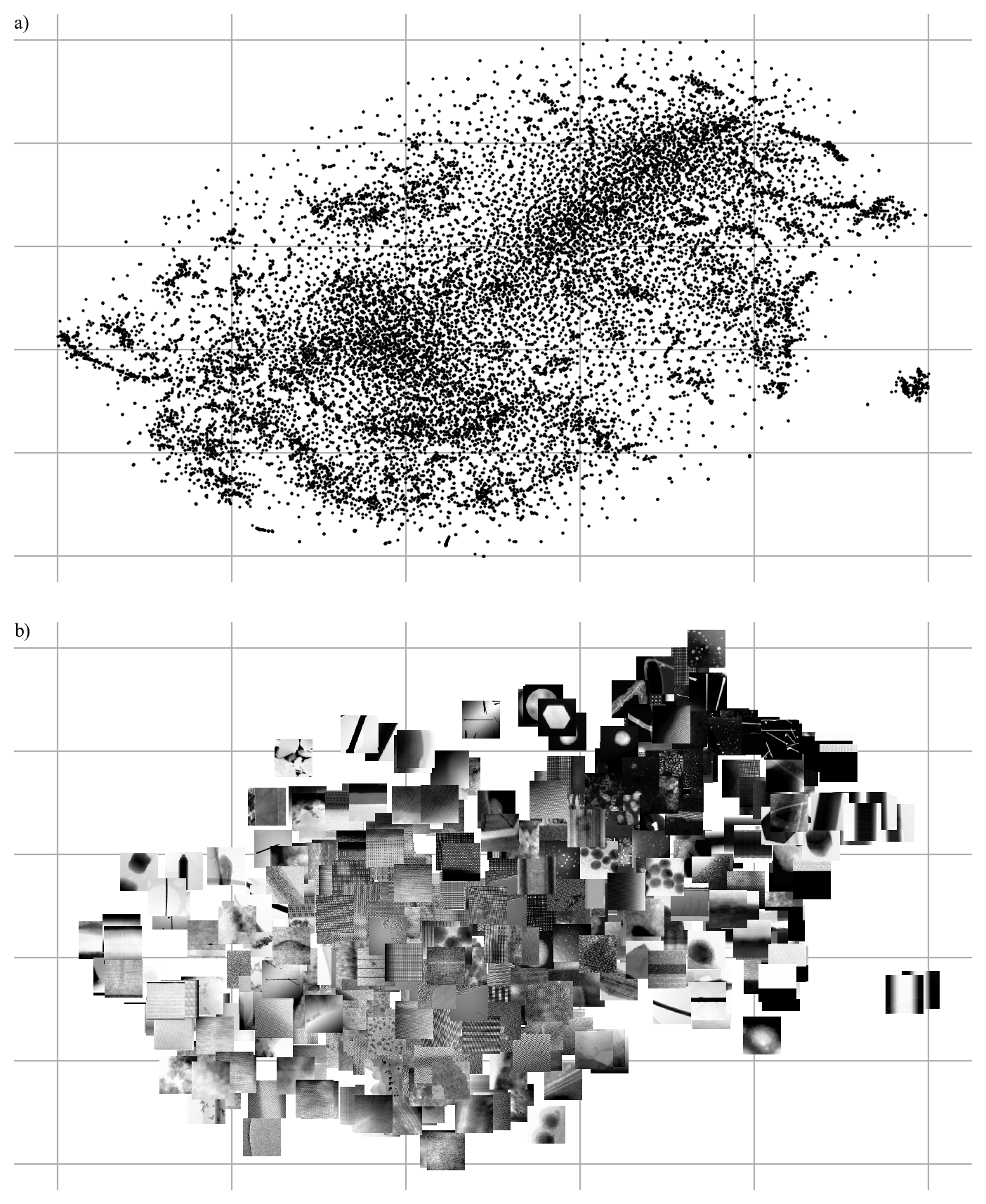}
\caption{ Two-dimensional tSNE visualization of 64-dimensional VAE latent spaces for 19769 STEM images that have been downsampled to 96$\times$96. The same grid is used to show a) map points and b) images at 500 randomly selected points. }
\label{stem_downsampled_96x96}
\end{figure*}

\begin{table}[tbh!]
\centering
\footnotesize
\begin{tabular*}{\textwidth}{c@{\extracolsep{\fill}}c}
\raisebox{-.5\height}{
\begin{overpic}[scale=\myimagescale]{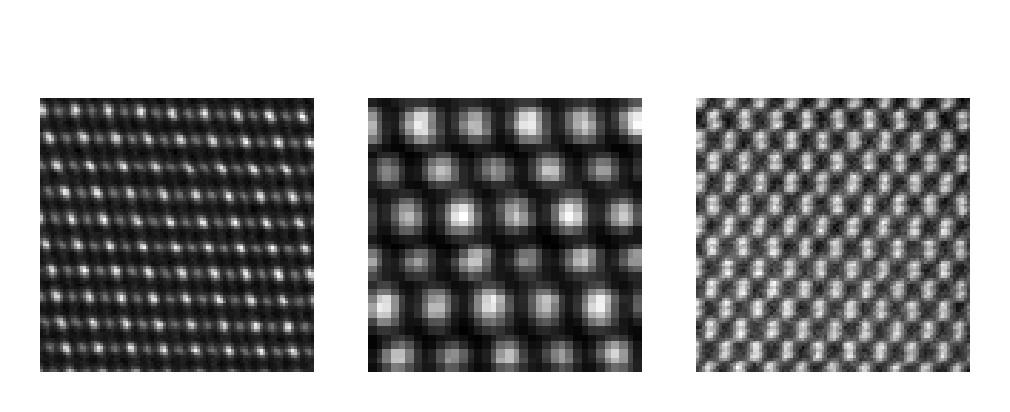}
\put(50,34){\makebox[0pt]{Dark Field Atom Columns \cite{van2016unscrambling}}}
\end{overpic}
} & \raisebox{-.5\height}{
\begin{overpic}[scale=\myimagescale]{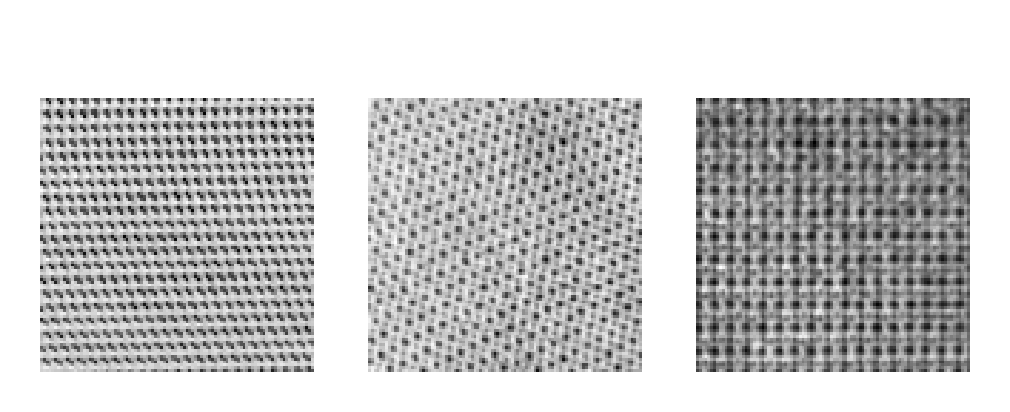}
\put(50,34){\makebox[0pt]{Bright Field Atom Columns \cite{zhou2016sample}}}
\end{overpic}
} \\
\raisebox{-.5\height}{
\begin{overpic}[scale=\myimagescale]{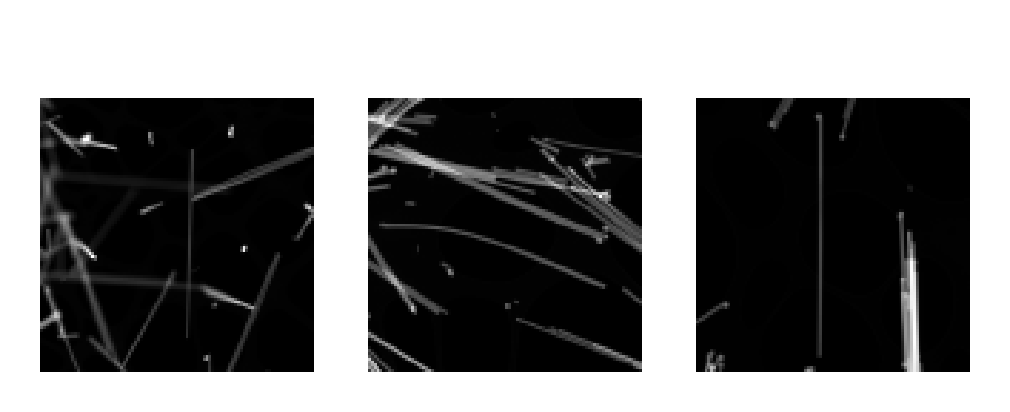}
\put(50,34){\makebox[0pt]{Nanowires \cite{bu2016surface}}}
\end{overpic}
} & \raisebox{-.5\height}{
\begin{overpic}[scale=\myimagescale]{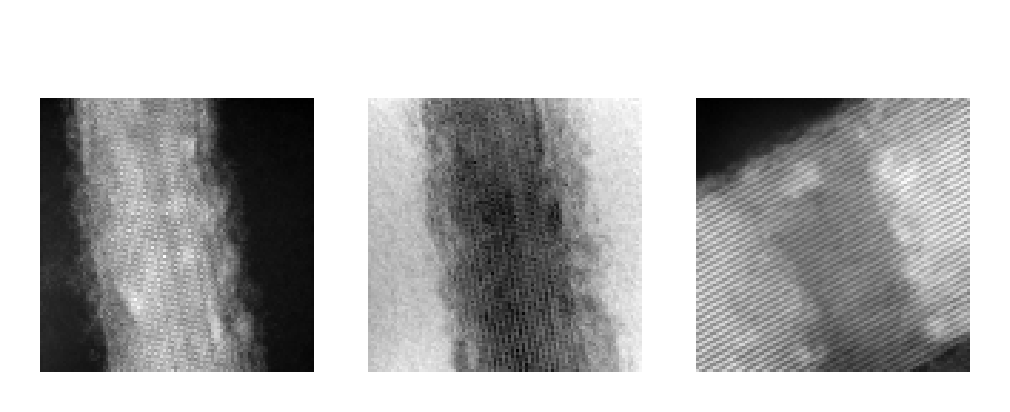}
\put(50,34){\makebox[0pt]{Atomic Resolution Bands}}
\end{overpic}
} \\
\raisebox{-.5\height}{
\begin{overpic}[scale=\myimagescale]{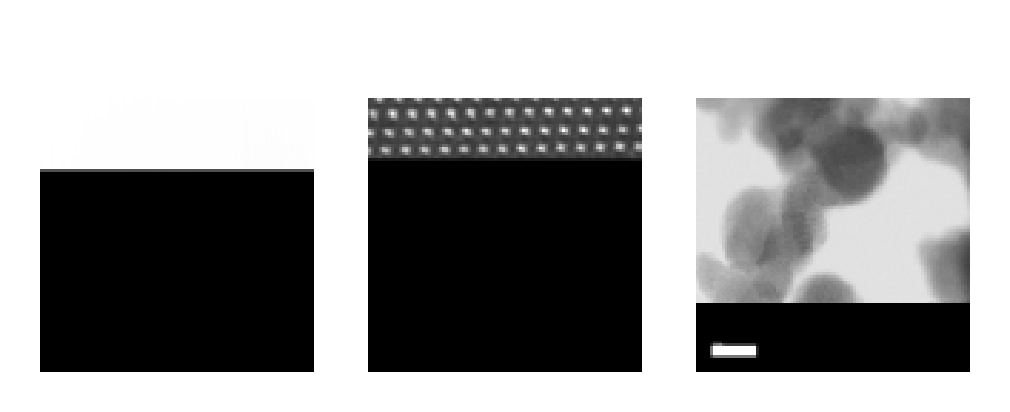}
\put(50,34){\makebox[0pt]{Incomplete Scans}}
\end{overpic}
} & \raisebox{-.5\height}{
\begin{overpic}[scale=\myimagescale]{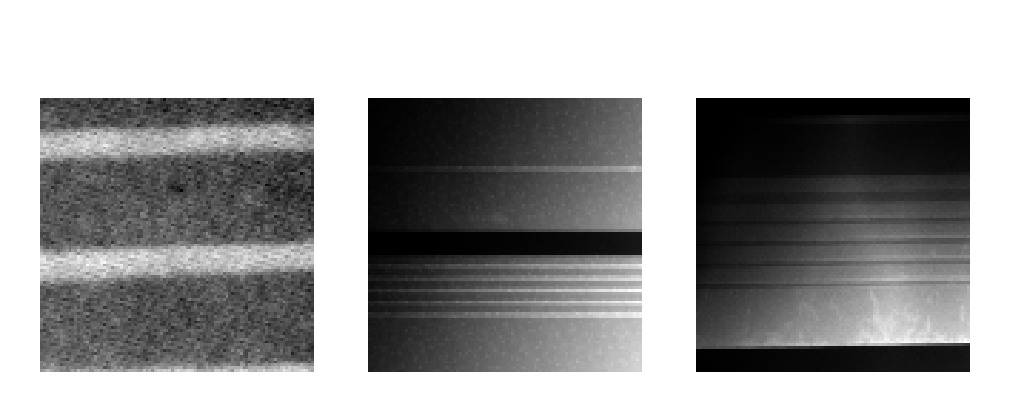}
\put(50,34){\makebox[0pt]{Multilayer Materials \cite{monclus2018effect}}}
\end{overpic}
} \\
\raisebox{-.5\height}{
\begin{overpic}[scale=\myimagescale]{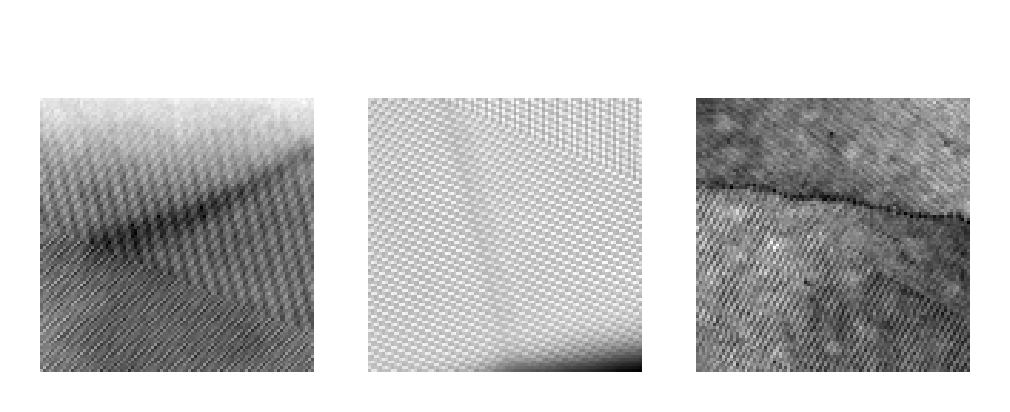}
\put(50,34){\makebox[0pt]{Atomic Boundaries \cite{pyrz2010atomic}}}
\end{overpic}
} & \raisebox{-.5\height}{
\begin{overpic}[scale=\myimagescale]{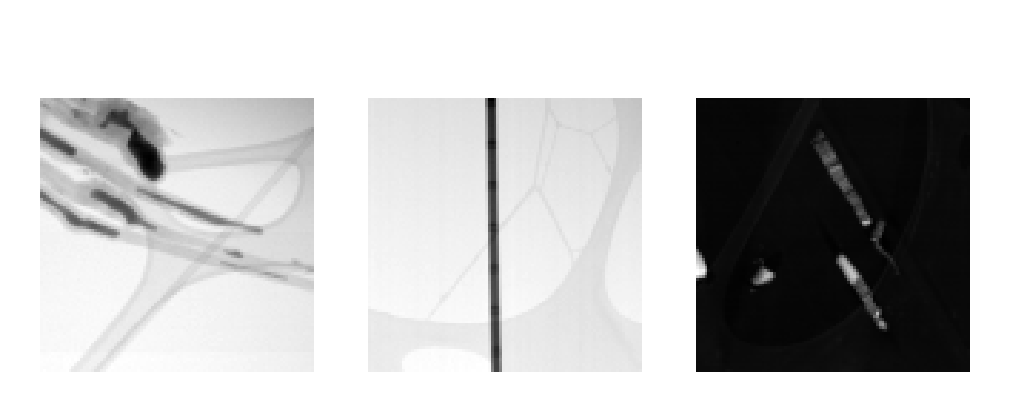}
\put(50,34){\makebox[0pt]{Lacey Carbon Supports \cite{mcgilvery2012contamination}}}
\end{overpic}
} \\
\end{tabular*}
\caption{ Examples and descriptions of STEM images in our datasets. References put some images into context to make them more tangible to unfamiliar readers. }
\label{table:stem_image_decriptions}
\end{table}

\section{Scanning Transmission Electron Micrographs}\label{sec:STEM}

We curated 19769 STEM images from University of Warwick electron microscopy dataservers to train ANNs for compressed sensing\cite{ede2020partial, preprint+ede2019deep}. Atom columns are visible in roughly two-thirds of images, and similar proportions are bright and dark field. In addition, most signals are noisy\cite{seki2018theoretical} and are imaged at several times their Nyquist rates\cite{landau1967sampling}. To reduce data transfer times for large images, we also created variant containing 161069 non-overlapping 512$\times$512 crops from full images. For rapid development, we have also created new variants containing 96$\times$96 images downsampled or cropped from full images. In this section we give details of each STEM dataset, referring to them using their names in our repositories.

\extravspace
\noindent \textbf{STEM Full Images:} 19769 32-bit TIFFs containing STEM images taken with a University of Warwick JEOL ARM 200F electron microscope by dozens of scientists working on hundreds of projects. Images were originally saved in DigitalMicrograph DM3 or DM4 files created by Gatan Microscopy Suite\cite{gms_webpage} software and have their original sizes and intensities. The dataset is partitioned into 14826 training, 1977 validation, and 2966 test set images. The dataset was made by concatenating contributions from different scientists, so partitioning the dataset before shuffling also partitions scientists.

\extravspace
\noindent \textbf{STEM Crops:} 161069 32-bit TIFFs containing 512$\times$512 non-overlapping regions cropped from STEM Full Images. The dataset is partitioned into 110933 training, 21259 validation, and 28877 test set images. This dataset is biased insofar that larger images were divided into more crops.

\extravspace
\noindent \textbf{STEM 96$\times$96:} A 32-bit NumPy\cite{npy_format, nep_npy_format} array with shape [19769, 96, 96, 1] containing 19769 STEM Full Images area downsampled to 96$\times$96 with MATLAB and default antialiasing.

\extravspace
\noindent \textbf{STEM 96$\times$96 Crops:} A 32-bit NumPy array with shape [19769, 96, 96, 1] containing 19769 96$\times$96 regions cropped from STEM Full Images. Each crop is from a different image.

Variety of STEM 96$\times$96 images is shown in figure~\ref{stem_downsampled_96x96} by clustering means and standard deviations of VAE latent spaces in two dimensions by tSNE. Details are in section~\ref{sec:visualization}. An interactive visualization that displays images when map points are hovered over is also available\cite{datasets_repo}. This paper is aimed at a general audience so readers may not be familiar with STEM. Subsequently, example images are tabulated with references and descriptions in table~\ref{table:stem_image_decriptions} to make them more tangible.

\begin{figure*}[tbp!]
\centering
\includegraphics[width=\textwidth]{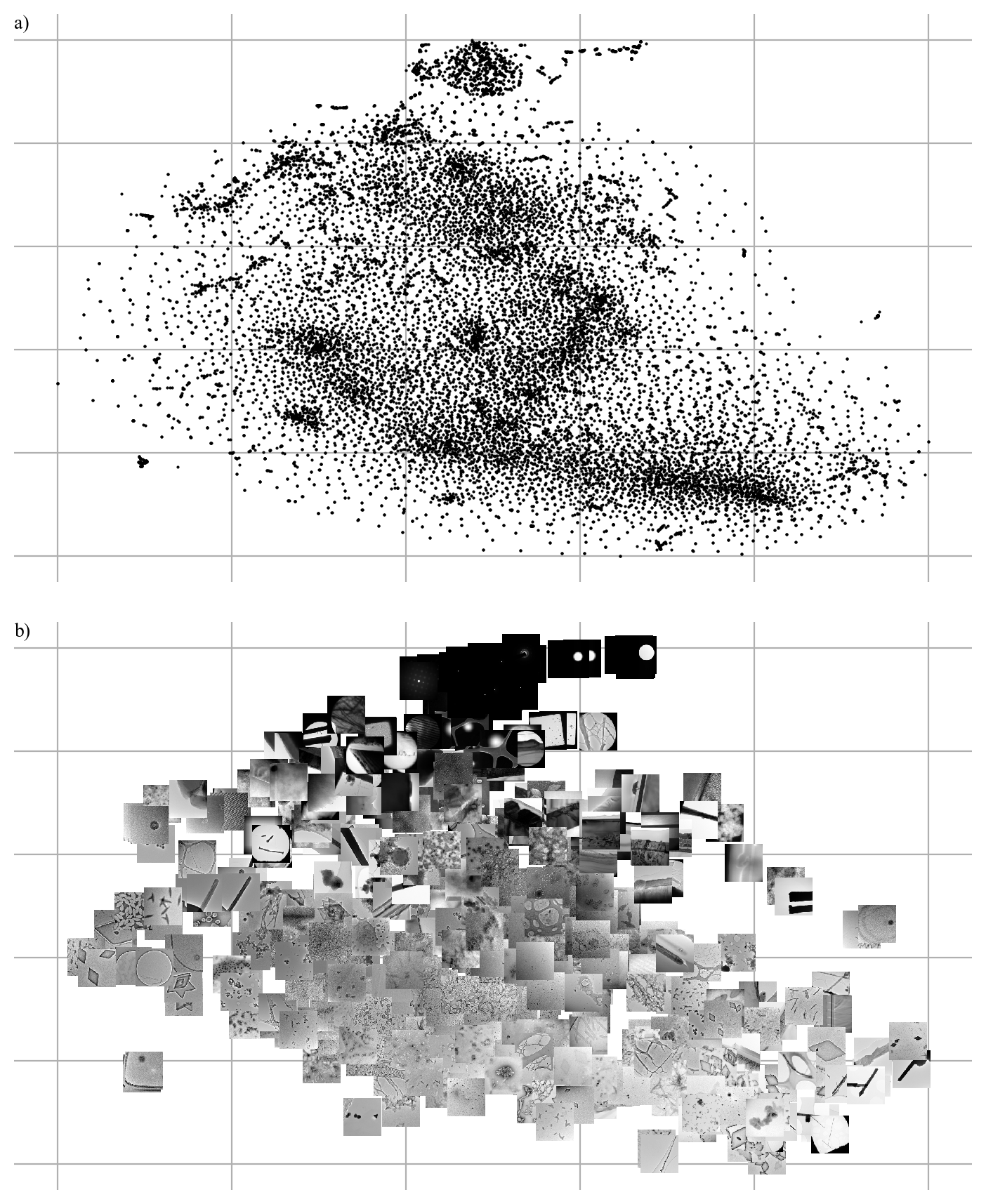}
\caption{ Two-dimensional tSNE visualization of 64-dimensional VAE latent spaces for 17266 TEM images that have been downsampled to 96$\times$96. The same grid is used to show a) map points and b) images at 500 randomly selected points. }
\label{tem_downsampled_96x96}
\end{figure*}

\begin{table}[tbh!]
\footnotesize
\centering
\begin{tabular*}{\textwidth}{c@{\extracolsep{\fill}}c}
\raisebox{-.5\height}{
\begin{overpic}[scale=\myimagescale]{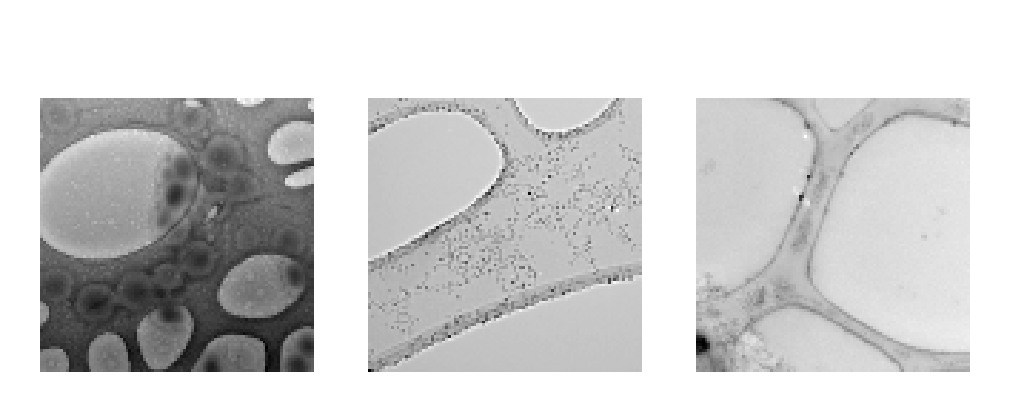}
\put(50,34){\makebox[0pt]{Lacey Carbon Supports \cite{karlsson2001thickness}}}
\end{overpic}
} & \raisebox{-.5\height}{
\begin{overpic}[scale=\myimagescale]{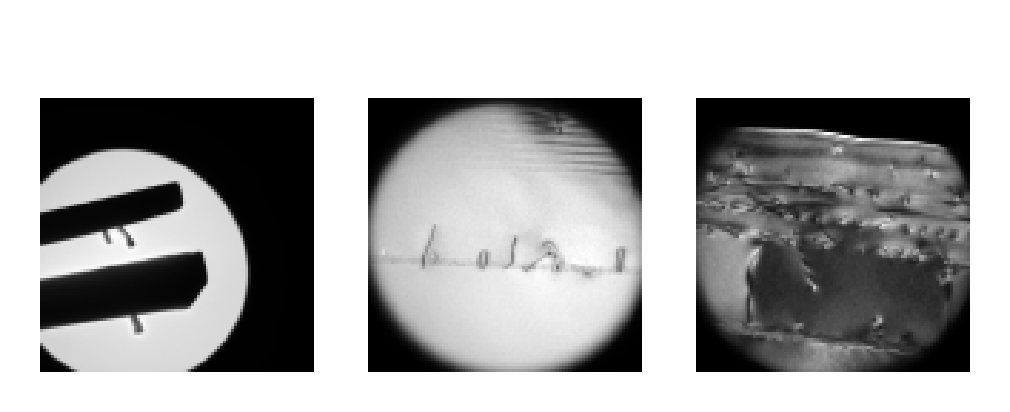}
\put(50,34){\makebox[0pt]{Apertures Blocking Electrons}}
\end{overpic}
} \\
\raisebox{-.5\height}{
\begin{overpic}[scale=\myimagescale]{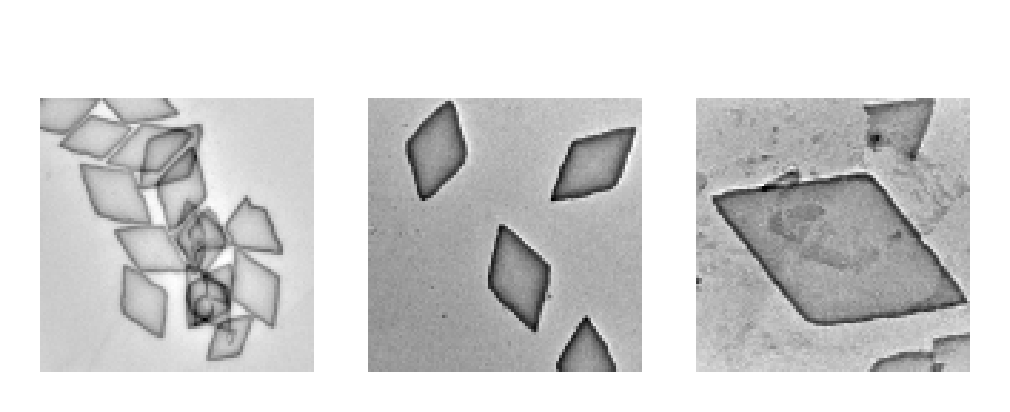}
\put(50,34){\makebox[0pt]{Block Copolymers \cite{inam20171d}}}
\end{overpic}
} & \raisebox{-.5\height}{
\begin{overpic}[scale=\myimagescale]{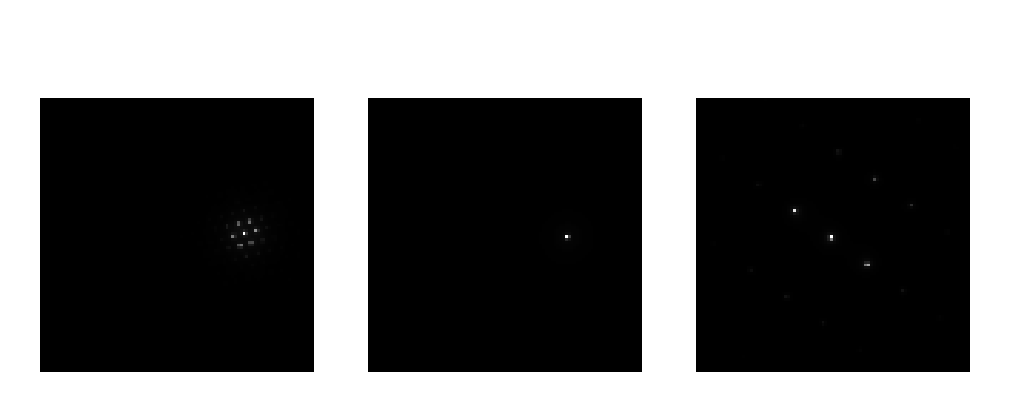}
\put(50,34){\makebox[0pt]{Diffraction Patterns \cite{bendersky2001electron}}}
\end{overpic}
} \\
\raisebox{-.5\height}{
\begin{overpic}[scale=\myimagescale]{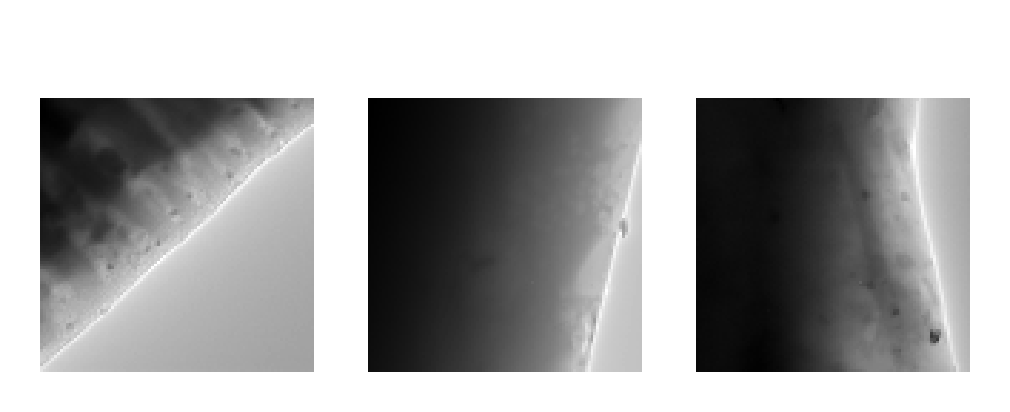}
\put(50,34){\makebox[0pt]{Vacuum at Specimen Edges}}
\end{overpic}
} & \raisebox{-.5\height}{
\begin{overpic}[scale=\myimagescale]{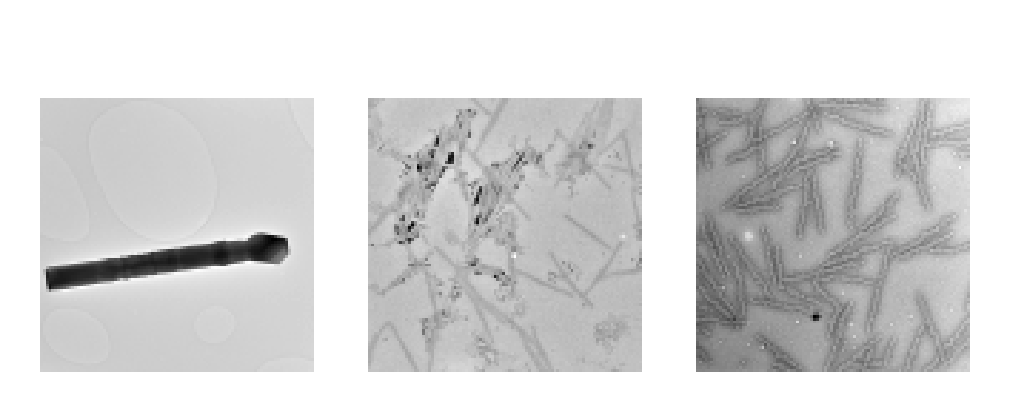}
\put(50,34){\makebox[0pt]{Nanowires \cite{wu2001superconducting}}}
\end{overpic}
} \\
\raisebox{-.5\height}{
\begin{overpic}[scale=\myimagescale]{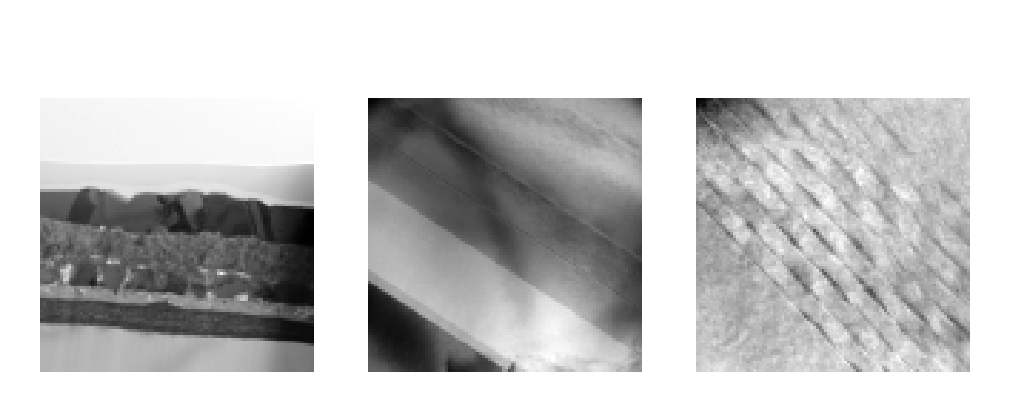}
\put(50,34){\makebox[0pt]{Multilayer Materials \cite{pang2017microstructural}}}
\end{overpic}
} & \raisebox{-.5\height}{
\begin{overpic}[scale=\myimagescale]{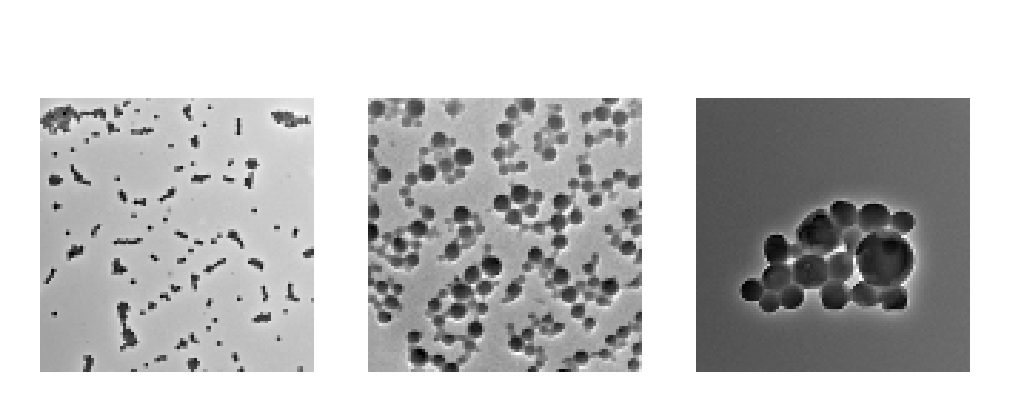}
\put(50,34){\makebox[0pt]{Particles \cite{dong2016individual}}}
\end{overpic}
} \\
\end{tabular*}
\caption{ Examples and descriptions of TEM images in our datasets. References put some images into context to make them more tangible to unfamiliar readers. }
\label{table:tem_image_decriptions}
\end{table}

\section{Transmission Electron Micrographs}

We curated 17266 2048$\times$2048 high-signal TEM images from University of Warwick electron microscopy dataservers to train ANNs to improve signal-to-noise\cite{ede2019improving}. However, our dataset was only available upon request. It is now openly available\cite{warwickem!}. For convenience, we have also created a new variant containing 96$\times$96 images that can be used for rapid ANN development. In this section we give details of each TEM dataset, referring to them using their names in our repositories.

\extravspace
\noindent \textbf{TEM Full Images:} 17266 32-bit TIFFs containing 2048$\times$2048 TEM images taken with University of Warwick JEOL 2000, JEOL 2100, JEOL 2100+, and JEOL ARM 200F electron microscope by dozens of scientists working on hundreds of projects. Images were originally saved in DigitalMicrograph DM3 or DM4 files created by Gatan Microscopy Suite\cite{gms_webpage} software and have been cropped to largest possible squares and area resized to 2048$\times$2048 with MATLAB and default antialiasing. Images with at least 2500 electron counts per pixel were then linearly transformed to have minimum and maximum values of 0 and 1, respectively. We discarded images with less than 2500 electron counts per pixel as images were curated to train an electron micrograph denoiser\cite{ede2019improving}. The dataset is partitioned into 11350 training, 2431 validation, and 3486 test set images. The dataset was made by concatenating contributions from different scientists, so each partition contains data collected by a different subset of scientists.

\extravspace
\noindent \textbf{TEM 96$\times$96:} A 32-bit NumPy array with shape [17266, 96, 96, 1] containing 17266 TEM Full Images area downsampled to 96$\times$96 with MATLAB and default antialiasing. Training, validation, and test set images are concatenated in that order. 

Variety of TEM 96$\times$96 images is shown in figure~\ref{tem_downsampled_96x96} by clustering means and standard deviations of VAE latent spaces in two dimensions by tSNE. Details are in section~\ref{sec:visualization}. An interactive visualization that displays images when map points are hovered over is also available\cite{datasets_repo}. This paper is aimed at a general audience so readers may not be familiar with TEM. Subsequently, example images are tabulated with references and descriptions in table~\ref{table:tem_image_decriptions} to make them more tangible.

\section{Exit Wavefunctions}

We simulated 98340 TEM exit wavefunctions to train ANNs to reconstruct phases from amplitudes\cite{preprint+ede2020exit}. Half of wavefunction information is undetected by conventional TEM as only the amplitude, and not the phase, of an image is recorded. Wavefunctions were simulated at 512$\times$512 then centre-cropped to 320$\times$320 to remove simulation edge artefacts. Wavefunctions have been simulated for real physics where Kirkland potentials\cite{kirkland2010advanced} for each atom are summed from $n=3$ terms, and by truncating Kirkland potential summations to $n=1$ to simulate an alternative universe where atoms have different potentials. Wavefunctions simulated for an alternate universe can be used to test ANN robustness to simulation physics. For rapid development, we also downsampled $n=3$ wavefunctions from 320$\times$320 to 96$\times$96. In this section we give details of each exit wavefunction dataset, referring to them using their names in our repositories.

\extravspace
\noindent \textbf{CIFs:} 12789 CIFs downloaded from the Crystallography Open Database\cite{Quiros2018, Merkys2016, Grazulis2015, Grazulis2012, Grazulis2009, Downs2003} (COD). The CIFs are for materials published in inorganic chemistry journals. There are 150 New Journal of Chemistry, 1034 American Mineralogist, 1998 Journal of the American Chemical Society and 5457 Inorganic Chemistry CIFs used to simulate training set wavefunctions, 1216 Physics and Chemistry of Materials CIFs used to simulate validation set wavefunctions, and 2927 Chemistry of Materials CIFs used to simulate test set wavefunctions. In addition, the CIFs have been preprocessed to be input to clTEM wavefunction simulations.

\extravspace
\noindent \textbf{URLs:} COD Uniform Resource Locators\cite{berners1994uniform} (URLs) that CIFs were downloaded from.

\extravspace
\noindent \textbf{Wavefunctions:} 36324 complex 64-bit NumPy files containing 320$\times$320 wavefunctions. The wavefunctions are for a large range of materials and physical hyperparameters. The dataset is partitioned into 24530 training, 3399 validation, and 8395 test set wavefunctions. Metadata Javascript Object Notation\cite{json} (JSON) files link wavefunctions to CIFs and contain some simulation hyperparameters.

\extravspace
\noindent \textbf{Wavefunctions Unseen Training:} 1544 64-bit NumPy files containing 320$\times$320 wavefunctions. The wavefunctions are for training set CIFs and are for a large range of materials and physical hyperparameters. Metadata JSONs link wavefunctions to CIFs and contain some simulation hyperparameters.

\extravspace
\noindent \textbf{Wavefunctions Single:} 4825 complex 64-bit NumPy files containing 320$\times$320 wavefunctions. The wavefunctions are for a single material, In$_{1.7}$K$_2$Se$_8$Sn$_{2.28}$\cite{hwang2004cooling}, and a large range of physical hyperparameters. The dataset is partitioned into 3861 training, and 964 validation set wavefunctions. Metadata JSONs link wavefunctions to CIFs and contain some simulation hyperparameters.

\extravspace
\noindent \textbf{Wavefunctions Restricted:} 11870 complex 64-bit NumPy files containing 320$\times$320 wavefunctions. The wavefunctions are for a large range of materials and a small range of physical hyperparameters. The dataset is partitioned into 8002 training, 1105 validation, and 2763 test set wavefunctions. Metadata JSON files link wavefunctions to CIFs and contain some simulation hyperparameters.

\extravspace
\noindent \textbf{Wavefunctions 96$\times$96:} A 32-bit NumPy array with shape [36324, 96, 96, 2] containing 36324 wavefunctions. The wavefunctions were simulated for a large range of materials and physical hyperparameters, and bilinearly downsampled with skimage\cite{van2014scikit} from 320$\times$320 to 96$\times$96 using default antialiasing. In Python\cite{python}, Real components are at index [...,0], and imaginary components are at index [...,1]. The dataset can be partitioned in 24530 training, 3399 validation, and 8395 test set wavefunctions, which have been concatenated in that order. To be clear, the training subset is at Python indexes [:24530].

\extravspace
\noindent \textbf{Wavefunctions 96$\times$96 Restricted:} A 32-bit NumPy array with shape [11870, 96, 96, 2] containing 11870 wavefunctions. The wavefunctions were simulated for a large range of materials and a small range of physical hyperparameters, and bilinearly downsampled with skimage from 320$\times$320 to 96$\times$96 using default antialiasing. The dataset can be partitioned in 8002 training, 1105 validation, and 2763 test set wavefunctions, which have been concatenated in that order.

\extravspace
\noindent \textbf{Wavefunctions 96$\times$96 Single:} A 32-bit NumPy array with shape [4825, 96, 96, 2] containing 11870 wavefunctions. The wavefunctions were simulated for In$_{1.7}$K$_2$Se$_8$Sn$_{2.28}$ and a large range of physical hyperparameters, and bilinearly downsampled with skimage from 320$\times$320 to 96$\times$96 using default antialiasing. The dataset can be partitioned in 3861 training, and 964 validation set wavefunctions, which have been concatenated in that order.

\extravspace
\noindent \textbf{Wavefunctions $n=1$:} 37457 complex 64-bit NumPy files containing 320$\times$320 wavefunctions. The wavefunctions are for a large range of materials and physical hyperparameters. The dataset is partitioned into 25352 training, 3569 validation, and 8563 test set wavefunctions. These wavefunctions are for an alternate universe where atoms have different potentials.

\extravspace
\noindent \textbf{Wavefunctions $n=1$ Unseen Training:} 1501 64-bit NumPy files containing 320$\times$320 wavefunctions. The wavefunctions are for training set CIFs and are for a large range of materials and physical hyperparameters. Metadata JSONs link wavefunctions to CIFs and contain some simulation hyperparameters. These wavefunctions are for an alternate universe where atoms have different potentials.

\extravspace
\noindent \textbf{Wavefunctions $n=1$ Single:} 4819 complex 64-bit NumPy files containing 320$\times$320 wavefunctions. The wavefunctions are for a single material, In$_{1.7}$K$_2$Se$_8$Sn$_{2.28}$, and a large range of physical hyperparameters. The dataset is partitioned into 3856 training, and 963 validation set wavefunctions. Metadata JSONs link wavefunctions to CIFs and contain some simulation hyperparameters. These wavefunctions are for an alternate universe where atoms have different potentials.

\extravspace
\noindent \textbf{Experimental Focal Series:} 1000 experimental focal series. Each series consists of 14 32-bit 512$\times$512 TEM images, area downsampled from 4096$\times$4096 with MATLAB and default antialiasing. The images are in TIFF\cite{adobe1992tiff} format. All series were created with a common, quadratically increasing\cite{haigh2013recording} defocus series. However, spatial scales vary and would need to be fitted as part of wavefunction reconstruction.

In detail, exit wavefunctions for a large range of physical hyperparameters were simulated with clTEM\cite{clTEM_repo, dyson2014advances} for acceleration voltages in $\{80, 200, 300\}$ kV, material depths uniformly distributed in $[5, 100)$ nm, material widths in $[5, 10)$ nm, and crystallographic zone axes $(h, k, l)$ $h, k, l \in \{0, 1, 2\}$. Materials were padded on all sides with vacuum 0.8 nm wide and 0.3 nm deep to reduce simulation artefacts. Finally, crystal tilts were perturbed by zero-centred Gaussian random variates with 0.1$\degree$ standard deviations. We used default values for other clTEM hyperparameters. Simulations for a small range of physical hyperparameters used lower upper bounds that reduced simulation hyperparameter ranges by factors close to 1/4. All wavefunctions are linearly transformed to have a mean amplitude of 1.

All wavefunctions show atom columns, so tSNE visualizations are provided in supplementary information to conserve space. The visualizations are for Wavefunctions 96$\times$96, Wavefunctions 96$\times$96 Restricted and Wavefunctions 96$\times$96 Single.

\section{Discussion}

The best dataset variant varies for different applications. Full-sized datasets can always be used as other dataset variants are derived from them. However, loading and processing full-sized examples may bottleneck training, and it is often unnecessary. Instead, smaller 512$\times$512 crops, which can be loaded more quickly the full-sized images, can often be used to train ANNs to be applied convolutionally\cite{zhu2017unpaired} to or tiled across\cite{ede2019improving} full-sized inputs. In addition, our 96$\times$96 datasets can be used for rapid initial development before scaling up to full-sized datasets, similar to how ANNs might be trained with CIFAR-10 before scaling up to ImageNet. However, subtle application- and dataset-specific considerations may also influence the best dataset choice. For example, an ANN trained with downsampled 96$\times$96 inputs may not generalize to 96$\times$96 crops from full-sized inputs as downsampling may introduce artifacts\cite{resize_artefacts} and change noise or other data characteristics. 

In practice, electron microscopists image most STEM and TEM signals at several times their Nyquist rates\cite{landau1967sampling}. This eases visual inspection, decreases sub-Nyquist aliasing\cite{amidror2015sub}, improves display on computer monitors, and is easier than carefully tuning sampling rates to capture the minimum data needed to resolve signals. High sampling may also reveal additional high-frequency information when images are inspected after an experiment. However, this complicates ANN development as it means that information per pixel is often higher in downsampled images. For example, partial scans across STEM images that have been dowsampled to 96$\times$96 require higher coverages than scans across 96$\times$96 crops for ANNs to learn to complete images with equal performance\cite{ede2020partial}. It also complicates the comparison of different approaches to compressed sensing. For example, we suggested that sampling 512$\times$512 crops at a regular grid of probing locations outperforms sampling along spiral paths as a subsampling grid can still access most information\cite{ede2020partial}.

Test set performance should be calculated for a standardized dataset partition to ease comparison with other methods. Nevertheless, training and validation partitions can be varied to investigate validation variance for partitions with different characteristics. Default training and validation sets for STEM and TEM datasets contain contributions from different scientists that have been concatenated or numbered in order, so new validation partitions can be selected by concatenating training and validation partitions and moving the window used to select the validation set. Similarly, exit wavefunctions were simulated with CIFs from different journals that were concatenated or numbered sequentially. There is leakage\cite{open2019how, bussola2019not} between training, validation and test sets due to overlap between materials published in different journals and between different scientists' work. However, further leakage can be minimized by selecting dataset partitions before any shuffling and, for wavefunctions, by ensuring that simulations for each journal are not split between partitions.

Experimental STEM and TEM image quality is variable. Images were taken by scientists with all levels of experience and TEM images were taken on multiple microscopes. This means that our datasets contain images that might be omitted from other datasets. For example, the tSNE visualization for STEM in figure~\ref{stem_downsampled_96x96} includes incomplete scans, $\sim$50 blank images, and images that only contain noise. Similarly, the tSNE visualization for TEM in figure~\ref{tem_downsampled_96x96} revealed some images where apertures block electrons, and that there are small number of unprocessed standard diffraction and convergent beam electron diffraction\cite{tanaka1994convergent} patterns. Although these conventionally low-quality images would not normally be published, they are important to ensure that ANNs are robust for live applications. In addition, inclusion of conventionally low-quality images may enable identification of this type of data. We encourage readers to try our interactive tSNE visualizations\cite{datasets_repo} for detailed inspection of our datasets. 

In this paper, we present tSNE visualizations of VAE latent spaces to show image variety. However, our VAEs can be directly applied to a wide range of additional applications. For example, successful tSNE clustering of latent spaces suggests that VAEs could be used to create a hash table\cite{pattersonsemantic, jin2019deep} for an electron micrograph search engine. VAEs can also be applied to semantic manipulation\cite{klys2018learning}, and clustering in tSNE visualizations may enable subsets of latent space that generate interesting subsets of data distributions to be identified. Other applications include using clusters in tSNE visualizations to label data for supervised learning, data compression, and anomaly detection\cite{yao2019unsupervised, xu2018unsupervised}. To encourage further development, we have made our source code and pretrained VAEs openly available\cite{datasets_repo}.

\section{Conclusion}

We have presented details of and visualizations for large new electron microscopy datasets that are openly available from our new repositories. Datasets have been carefully partitioned into training, validation, and test sets for machine learning. Further, we provide variants containing 512$\times$512 crops to reduce data loading times, and examples downsampled to 96$\times$96 for rapid development. To improve dataset visualization with VAEs, we introduce encoding normalization and regularization, and add an image gradient loss. In addition, we propose extending tSNE to account for encoded standard deviations. Source code, pretrained VAEs, precompiled tSNE binaries, and interactive dataset visualizations are provided in supplementary repositories to help users become familiar with our datasets and visualizations. By making our datasets available, we aim to encourage standardization of performance benchmarks in electron microscopy and increase participation of the wider computer science community in electron microscopy research. 

\section{Supplementary Information}

\textit{Note: Content in this section will be provided in a supplementary document with the published version of this preprint.}

\noindent Ten additional tSNE visualizations are provided as supplementary information. They are for:
\begin{itemize}
    \item Extracting 50 principal components by probabilistic PCA for the STEM 96$\times$96, STEM 96$\times$96 Crops, TEM 96$\times$96, Wavefunctions 96$\times$96, Wavefunctions 96$\times$96 Restricted and Wavefunctions 96$\times$96 Single datasets. PCA is a quick and effective method to extract features. As a result, we think that visualizations for PCA are interesting benchmarks.
    \item VAE latent spaces with $\boldsymbol\sigma$ propagation for the STEM 96$\times$96 Crops dataset. Crops show smaller features than downsampled images.
    \item VAE latent spaces without $\boldsymbol\sigma$ propagation for the STEM 96$\times$96, STEM 96$\times$96 Crops and TEM 96$\times$96 datasets. They are comparable to visualizations created with $\boldsymbol\sigma$ propagation.
\end{itemize}
Interactive versions of tSNE visualizations that display data when map points are hovered over are available\cite{datasets_repo} for every figure. In addition, we propose an algorithm to increase whitespace utilization in tSNE visualizations by uniformly separating points. 

\section{Data Availability}

The data that support the findings of this study are openly available at \url{https://doi.org/10.5281/zenodo.3834197}. For additional information contact the corresponding author (J.M.E.).

\bibliography{bibliography}

\section{Acknowledgements}

Funding: J.M.E. acknowledges EPSRC EP/N035437/1 and EPSRC Studentship 1917382.

\section{Competing Interests}

The author declares no competing interests.


\end{document}